\documentclass[letter,twocolumn]{jpsj2}

\title{%
Quasiparticle Heat Transport in Mixed State of High-$T_c$ Superconductors
}

\author{%
Mitsuaki Takigawa\thanks{E-mail address: takigawa@mp.okayama-u.ac.jp}, 
Masanori Ichioka and 
Kazushige Machida
}

\inst{%
Department of Physics, Okayama University, Okayama 700-8530, Japan
}

\recdate{\today}

\abst{%
The field dependence of low-temperature thermal conductivity  $\kappa(H)$
observed on cuprates is explained by calculating $\kappa(H)$ 
microscopically. 
The heat current carried by low-lying quasiparticles 
around a vortex core decreases with $H$ due to its depletion by the 
$H$-induced moment centered at a core for an underdoped case.
This leads to a common and consistent picture on an ``anomalous core''.
$\kappa(H,T)$ is found to be a useful tool for probing quasiparticle
structures in the mixed state in general.
}

\kword{%
thermal conductivity, vortex lattice state, antiferromagnetism, 
Bogoliubov-de Gennes theory, extended Hubbard model
}

\begin{document}
\maketitle

%%%%%%%%%%%%%  INTRODUCTION
Much attention has been focused on high-temperature 
cuprate superconductors since their discovery over 15 years ago.
Although no consensus has been reached yet as to the
pairing mechanism to induce high $T_c$, we are gaining deep basic
knowledge regarding unconventional $d$-wave superconductivity
in general.
Recently, a series of experimental and theoretical studies 
have revealed significant phenomena associated with a quantized vortex in 
high-$T_c$ superconductors. A mere few Tesla field has discovered hidden
superconducting ground state properties 
that could not be observed by conventional 
bulk measurements. Namely, a magnetic field applied to the $c$-axis
induces magnetic moments exclusively around a vortex core.
This is evidenced by neutron scattering 
experiments~\cite{yamada,lake1,lake2,khaykovich} which
show an enhancement of the static incommensurately modulated 
moments under a field in La$_{2-x}$Sr$_{x}$CuO$_{4+\delta}$ (LSCO). 
A muon spin resonance experiment~\cite{sonier} 
also shows an enhanced magnetic activity under a field.  
The antiferromagnetism (AFM)-induced vortex state is 
reproduced by some theoretical works~\cite{AFM1,AFM2,AFM3,AFM4,takigawa2}.

In addition to these experiments 
which probe the moment increment averaged 
over an entire sample, other local probes reveal 
that magnetic moments exclusively appear locally 
around a vortex core. A scanning tunneling microscopy (STM) experiment~\cite{hoffman} 
on Bi2212 directly reveals a checkerboard pattern 
of the local density of states (LDOS) with a four-atomic-site period 
around a core, which strongly matches the neutron results with 
eight-site spin modulation. 
Site-selective NMR experiments on YBCO~\cite{halperin} and 
Tl$_2$Ba$_2$CuO$_{6+x}$~\cite{kakuyanagi1,kakuyanagi2} 
also demonstrate that the relaxation time at the core site 
becomes longer than those at other sites away from the core, 
implying that LDOS at the core site decreases with $H$.
These experiments collectively point to the ``anomalous core'' in cuprates. 
This picture allows us to understand 
a long standing puzzle (``empty core'') of the absence 
of the zero-energy DOS (ZEDOS) peak expected 
for a $d$-wave pairing vortex as observed by STM~\cite{maggio,renner,pan}.

Recent thermal conductivity measurements~\cite{sun,hawthorn} on LSCO under $H$
are interesting in this respect because heat current at low $T$ is
carried by quasiparticles associated with vortices. 
Thus, thermal conductivity provides yet another important information 
concerning LDOS around a vortex. 
Two independent studies 
by Sun \textit{et al.}~\cite{sun} and Hawthorn 
\textit{et al.}~\cite{hawthorn} yield essentially the same result: 
According to Sun \textit{et al.} 
(1) for an overdoped region ($x=0.17$ and $x=0.22$), 
thermal conductivity $\kappa(H)/\kappa(0)$ increases 
as a function of $H$ at low $T$ while it decreases as $T$ increases. 
At an intermediate temperature 
$T^{\ast}$, $\kappa(H)/\kappa(0)$ becomes almost constant 
as $H$ increases (Fig. 2 in ref. \citen{sun}).
(2) For a sample with an intermediate 
doping $x=0.14$, $\kappa(H)/\kappa(0)$ decreases at low $T$. 
This decreasing tendency is strengthened 
as $T$ increases (Fig. 3(c) in ref. \citen{sun}).
(3) For underdopings with $x=0.10$ and $x=0.08$, 
$\kappa(H)/\kappa(0)$ under a fixed field is minimum 
at a finite temperature (see Figs. 3(a) and 3(b) in ref. \citen{sun}).
(We define this characteristic temperature as $T_{\rm min}$.)
The essential features of these results are also 
reproduced by Hawthorn \textit{et al.}~\cite{hawthorn}; 
in particular, at the lowest $T$(= 60 mK), $\kappa(H)/\kappa(0)$ 
decreases (increases) for underdoped (overdoped) 
samples (Fig. 2 in ref. \citen{hawthorn}). 
They attribute this behavior to 
``field-induced thermal metal-to-insulator transition''.
Similar behaviors are also observed in YBCO.~\cite{sun2}

Here, we are going to explain these seemingly complex,
yet rich  behaviors by calculating $\kappa(H,T)$ microscopically
in the mixed state for $d$-wave superconductors.~\cite{gusynin}
We assume the possibility that underdoped 
cuprates have an intrinsic tendency towards 
AFM instability. 
This assumption is completely justifiable 
both experimentally as mentioned above and physically
because it turns out that the induced 
moment appears exclusively at the core sites 
where superconductivity is locally weakened, 
revealing a ``true'' ground state when 
removing superconductivity~\cite{takigawa2}.

Before going into detailed computations, 
we provide a clear view for understanding these phenomena. 
In the overdoped region away from magnetism, 
a standard $d$-wave vortex picture must be valid.
$\kappa(H)$ is basically an increasing function of $H$ at low $T$
because of the growth of ZEDOS.
At high $T$ more than $T^{\ast}$, $\kappa(H)$ decreases 
because of the inability of the ZEDOS 
at the core to contribute to $\kappa$ in high $T$ 
whose region increases with $H$ 
(see the detailed exposition in ref. \citen{takigawa}).
Towards underdopings, the system exhibits AFM 
instability exclusively at the vortex core, 
leading to the removal of ZEDOS. 
Thus, $\kappa(H)$ decreases with $H$. 
This instability is further amplified as $H$ grows 
since the AFM moment is enhanced.

%%%%%%%%%%%%%  FORMULATION
We start with an extended Hubbard model on a two-dimensional 
square lattice, and introduce the mean fields 
$n_{i,\sigma}= \langle a^\dagger_{i,\sigma} a_{i,\sigma} \rangle$
at the $i$-site, where $\sigma$ is the spin index and $i=(i_x,i_y)$ and 
$\Delta_{{\hat e},i,\sigma}=\langle a_{i,-\sigma} a_{i+\hat{e},\sigma} \rangle$.
We assume a pairing interaction $V$ between nearest-neighbor (NN) sites. 
This type of pairing interaction gives $d$-wave 
superconductivity.
Thus, the mean-field Hamiltonian under $H$~\cite{takigawa2,oka2001}
is given by 
%%
%\begin{eqnarray}
\begin{align}
{\cal H}=
&-\sum_{i,j,\sigma} \tilde{t}_{i,j}a^{\dagger}_{i,\sigma} a_{j,\sigma}
+U\sum_{i,\sigma} n_{i,-\sigma}
a^\dagger_{i,\sigma} a_{i,\sigma} 
\nonumber \\ 
&+V\sum_{\hat{e},i,\sigma}
(\Delta^{*}_{\hat{e},i,\sigma} a_{i,-\sigma} a_{i+\hat{e},\sigma}
+\Delta_{\hat{e},i,\sigma} a^\dagger_{i,\sigma} a^\dagger_{i+\hat{e},-\sigma}
),\qquad
%\label{eq:2.2}
\end{align}
%\end{eqnarray}
where $a^{\dagger}_{i,\sigma}$
($a_{i,\sigma}$) is the creation (annihilation) operator, and 
$i+\hat{e}$ represents the NN site ($\hat{e}=\pm\hat{x},\pm\hat{y}$).
The transfer integral expressed as 
$\tilde{t}_{i,j}$ is modified by the external field $H$ through the 
Peierls phase factor. The original hopping integral $t_{i,j}$ defined 
on a square lattice is assumed to be $t_{i,j}=t$ for the first NN pair, 
$t'=-0.12t$ for the second NN pair and $t''=0.08t$ for the third NN pair, 
which are selected to reproduce the Fermi surface topology in LSCO~\cite{tohyama}.
The average electron density per site is fixed at $\sim 0.875$.
We studied the square vortex lattice with the unit cell size $N_r\times N_r$
where two vortices are accommodated. $H=2\phi_0/a^2N_r^2$ with $a$
the atomic lattice constant and $\phi_0$ the unit flux. 
The field strength is denoted by $H_{N_r}(=1/N_r^2)$.

The large AFM moment at underdoping decreases with 
increasing hole doping, and vanishes at overdoping 
due to the absence of the nesting of the Fermi surface.
This magnetic tendency is simulated by changing $U$ 
in our calculation 
since the following results dominantly 
depend on the amplitude of AFM, 
rather than the Fermi surface shape.
We select $U/t=0.0$,  $U/t=2.6$ and $U/t=3.0$, and set $V/t=-2.0$. 
These $U$ values are designed to mimic the behaviors of overdoped, 
intermediate and underdoped cases, respectively, in LSCO.
For $U=0$, $\Delta_0=1.0t$, and $T_c$ $\sim$ $0.41t$.

To determine the eigenenergy $E_{\alpha}$ and the wave functions
$u_\alpha(\mib{r}_i)$ and $v_\alpha(\mib{r}_i)$ at the $i$-site,
we solve the Bogoliubov-de Gennes equation given by
%%%
%\begin{eqnarray}
\begin{align}
\sum_j
\left( \begin{array}{cc}
K_{\uparrow,i,j} & D_{i,j} \\ D^\dagger_{i,j} & -K^\ast_{\downarrow,i,j}
\end{array} \right)
\left( \begin{array}{c} u_\alpha(\mib{r}_j) \\ v_\alpha(\mib{r}_j)
\end{array}\right)
=E_\alpha
\left( \begin{array}{c} u_\alpha(\mib{r}_i) \\ v_\alpha(\mib{r}_i)
\end{array}\right),
\label{eq:BdG1}
\end{align}
%\end{eqnarray}
%%%
where
$K_{\sigma,i,j}=-\tilde{t}_{i,j} +\delta_{i,j} (U n_{i,-\sigma} -\mu)$ 
with the chemical potential $\mu$, 
$D_{i,j}=V \sum_{\hat{e}}  \Delta_{i,j} \delta_{j,i+\hat{e}} $ 
and $\alpha$ is the index of the eigenstate.
The self-consistent condition for the pair potential is
$\Delta_{i,j}=-\frac{1}{2}\sum_\alpha u_\alpha(\mib{r}_i)
v^\ast_\alpha(\mib{r}_j) \tanh(E_\alpha /2T).$
The so-called Dopper shift effect is also included through phase winding 
of $\Delta$ around a vortex.
The charge densities are 
$n_{i,\uparrow}=\sum_\alpha |u_\alpha(\mib{r}_i)|^2 f(E_\alpha )$ and
$n_{i,\downarrow}=\sum_\alpha |v_\alpha(\mib{r}_i)|^2 (1-f(E_\alpha ))$.

We calculate thermal conductivity $\kappa$ following the standard 
linear response theory. 
All the details are described in our previous paper 
studying the $U=0$ case~\cite{takigawa}.
To recapitulate briefly, the heat current flows 
in response to a small temperature gradient. 
The local thermal conductivity $\kappa(\mib{r})$ is written as 
%\begin{eqnarray} && 
\begin{align}
\kappa_{xx}(\mib{r})
&=\frac{h_x(\mib{r})}{-\nabla_xT} 
\nonumber \\ 
&=\frac{1}{T}{\rm Im}\{
\frac{\rm d}{{\rm d}\Omega} \frac{1}{N} \sum_{\mib{r}'} Q_{xx}(\mib{r}, 
\mib{r}', {\rm i}\Omega_n\rightarrow\Omega+{\rm i}0^+ )\}, 
\label{eq:kappar} 
%\end{eqnarray} 
\end{align}
where the heat-current correlation function $Q_{xx}$ is expressed 
by the above eigenvalues and eigenfunctions.
The spatially averaged thermal conductivity 
%\begin{equation} 
$\kappa_{xx}=\frac{1}{N}\sum_{\mib{r}}\kappa_{xx}(\mib{r})$ 
%\end{equation} 
is observed experimentally.

%%%%%%%%%%%%%%%%%%%%%%%%%%%%%
\begin{figure*}[tbh]
\includegraphics[width=16cm,height=5cm]{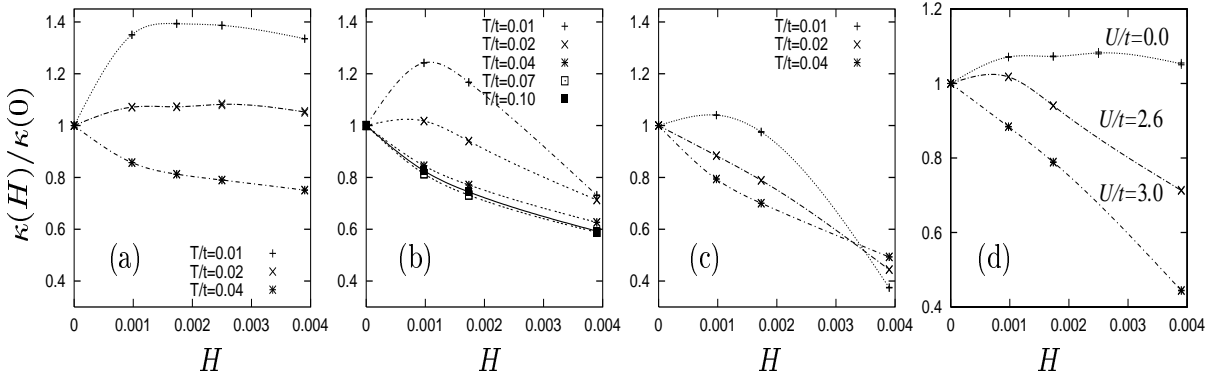}% Here is how to import EPS art
\caption{\label{fig:K-H}
Field dependence of $\kappa(H)/\kappa(0)$.
 $U/t=0.0$ (a), $2.6$ (b), $3.0$ (c) and
(d) $T=0.02t$ for $U/t=0.0$, $2.6$ and $3.0$.
The horizontal axis denotes $H_{N_r}=1/N^2_r$ ($N_r$ the magnetic unit cell size).
Note that $T_{\rm min} \sim 0.06t$ in (b).
}
\end{figure*}
%%%%%%%%%%%%%%%%%%%%%%%%%%%%%

In Fig. \ref{fig:K-H}, we show our results of $\kappa(H)/\kappa(0)$
as a function of $H$ for the selected temperatures at $U=0$ (a), 
$U/t=2.6$ (b) and $U/t=3.0$ (c). 
%As noted in Fig. \ref{fig:K-H}(a) in agreement with 
%the overdoping case that at the lowest $T$, 
It is seen from Fig. \ref{fig:K-H}(a) corresponding to the overdoping case 
that at the lowest $T$, 
$\kappa(H)$ increases at low $H$ corresponding to 
the fact that ZEDOS at the core grows with $H$.
At high $T$ ($T>T^{\ast}$), $\kappa(H)$ decreases 
with $H$ (with increasing vortex number) 
because the heat flow $\kappa(\mib{r})$ 
is suppressed at the core~\cite{takigawa}.
Near $T\sim T^{\ast}$, $\kappa(H)$ becomes almost independent of $H$, 
exhibiting a plateau behavior. 
These behaviors are seen in the data $x=0.17$ and $x=0.22$ 
obtained by Sun \textit{et al.}~\cite{sun}
where the corresponding $T^{\ast}$'s are $\sim$ 3 K and 5 K, 
respectively (see Fig. 2 in ref. ~\citen{sun}). 
Various $\kappa(T)$ curves for different 
fields pass through $T^{\ast}$~\cite{ando}. 
This focal point temperature $T^{\ast}$ is determined 
by the energy width of the zero-energy peak at the core.
This plateau behavior of $\kappa(H)$ 
%agrees with 
is reminiscent of the data 
obtained by Krishana \textit{et al.}~\cite{krishana} 
who attributed this to the field-induced transition 
from a $d_{x^2-y^2}$-wave pairing to a pairing with a full gap 
such as $d_{x^2-y^2}$+i$d_{xy}$ or $d_{x^2-y^2}$+{\rm i}$s$.

The most intricate and interesting case is shown in Fig. \ref{fig:K-H}(b) 
where we depict the results in $U/t=2.6$ 
corresponding to intermediate dopings. 
At the lowest $T$, $\kappa(H)$ initially increases at low $H$. 
Then upon increasing $T$, $\kappa(H)$ begins to decrease monotonically, 
and therefore further increasing $T$, 
the initial suppression decreases. 
Thus, the overall curve $\kappa(H)$
is pushed up again at $T=T_{\rm min}$ ($\sim 0.06t$ when $U/t=2.6$). 
This ``reentrant'' behavior is precisely reproduced 
experimentally by Sun \textit{et al.}~\cite{sun} 
(Figs. 3(a) and 3(b) in ref. \citen{sun} for $x=0.08$ and 0.10). 
It is noted, however, that the peak behaviors at low temperatures 
(Figs. 1(b) and 1(c)) are not noted in the experiment.
In these cases, $T_{\rm min}$ is 3-5K. 
We note here that $T_{\rm min}$ decreases when $U$ 
increases as explained later in Fig. 3.
It is understandable why there is no ``reentrant" behavior observed 
in the $x=0.14$ data (Fig. 3(c) in ref. ~\citen{sun}) 
because $T_{\rm min}$ does not reach up to $T<7$K in this experiment.
In the $U/t=3.0$ case shown in Fig. \ref{fig:K-H}(c), 
where at a finite temperature, 
the local AFM order centered around a vortex core appears, 
$\kappa(H)$ strongly decreases with $H$.

In Fig. \ref{fig:K-H}(d), we display the field dependences of $\kappa(H)/\kappa(0)$ 
for three $U$ values  at a fixed temperature. 
Depending on $U$, $\kappa(H)$ decreases or increases at low $H$. 
These contrasting behaviors correspond to the principal 
observation by Hawthorn \textit{et al}~\cite{hawthorn}. 
Namely, for the underdoped samples ($x=0.06$ and 0.09), $d\kappa(H)/dH<0$, 
while for $x=0.17$ and 0.20, $d\kappa(H)/dH>0$.
This observation is basically consistent 
with that by Sun \textit{et al.}~\cite{sun} 
where the increasing and decreasing tendencies in $\kappa(H)$ 
is divided at the optimal doping (Fig. 4 in ref. ~\citen{sun}).
Hawthorn \textit{et al.} interpreted this as coming from 
the field-induced thermal metal-insulator transition 
because the underdoped systems exhibit 
lack of electronic heat conduction. 
Here, we succeed in reproducing this ``transition'' 
by the introduction of the AFM local order in the underdoped case, 
which effectively removes the quasiparticles around a core, 
leading to poor thermal conductivity.

%%%%%%%%%%%%%%%%%%%%%%%%%%%%%
\begin{figure}[bth]
\includegraphics[width=7cm]{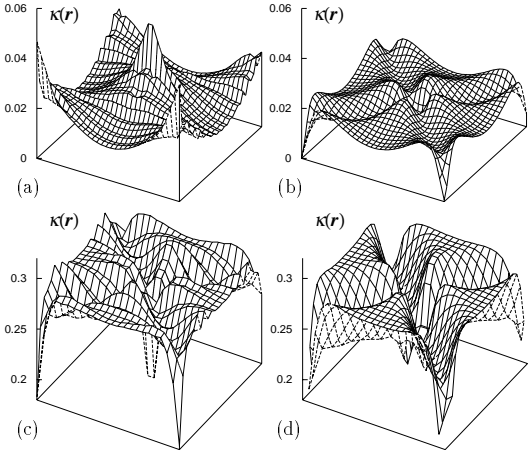}% Here is how to import EPS art
\caption{\label{fig:K-r}
Spatial structure of 
effective local thermal conductivity $\kappa(\mib{r})$
at $H_{N_r}=H_{32}$ at $T=0.01t$ [(a) and (b)] 
and at $T_{\rm min}$ [(c) and (d)], 
when $U=0$ [(a) and (c)], and $U=3.0$ [(b) and (d)].
The vortex cores are located at a corner and the center 
of the unit cell}
\end{figure}
%%%%%%%%%%%%%%%%%%%%%%%%%%%%%

To understand this field-induced transition more closely, 
we analyze the spatially dependent thermal conductivity $\kappa(\mib{r})$, 
indicating where the heat current is mainly carried spatially 
in the vortex lattice. 
At low temperature, $\kappa(H)$ reflects the zero-energy LDOS~\cite{takigawa}.
Therefore, when $U=0$ [Fig. \ref{fig:K-r}(a)], $\kappa(\mib{r})$ has a peak 
at the vortex core, where the ZEDOS peak appears.
However, when AFM appears  and ZEDOS is suppressed 
at the vortex core, $\kappa(\mib{r})$ is also 
suppressed at the core [Fig. \ref{fig:K-r}(b)].
Roughly speaking, $\kappa(\mib{r_c})$ at the farthest site $\mib{r_c}$ 
from vortices is nearly equal to $\kappa(H=0)$.
Therefore, since the spatial average of $\kappa(\mib{r})$ is 
larger (smaller) than $\kappa(\mib{r_c})$, 
we obtain $\kappa(H)/\kappa(0)>0(<0)$ 
in the case of Fig. \ref{fig:K-r}(a)(Fig. \ref{fig:K-r}(b)).
In Figs. \ref{fig:K-r}(c) and \ref{fig:K-r}(d), 
we show $\kappa(\mib{r})$ at $T_{\rm min} \sim 0.07t$.
In this high-temperature case, since the contribution to $\kappa(\mib{r})$ 
comes from the LDOS at energys near $\Delta_0$, $\kappa(\mib{r})$ is suppressed 
at the core~\cite{takigawa}. 
That is, the vortex core pushes aside the thermal flow.
Therefore, the spatial average of $\kappa(\mib{r})$ 
is smaller than $\kappa(\mib{r_c})$, 
resulting in $\kappa(H)<\kappa(0)$ at a high temperature.

In the presence of AFM, the LDOS is different 
between up-spin electrons ($N_\uparrow(\mib{r},E)$) and 
down-spin electrons ($N_\downarrow(\mib{r},E)$), 
depending on the sign of the AFM moment.
When $N_\uparrow(\mib{r},E)>N_\downarrow(\mib{r},E)$ at a site, 
$N_\uparrow(\mib{r},E)<N_\downarrow(\mib{r},E)$ at the nearest site.
As the electron spin is conserved in the electron hopping, 
$\kappa(\mib{r})$ is associated with the smaller LDOS 
$\sqrt{N_\uparrow(\mib{r},E) N_\downarrow(\mib{r},E)}$
, rather than 
$(N_\uparrow(\mib{r},E) + N_\downarrow(\mib{r},E))/2$.
With increasing $U$, the difference between 
$N_\uparrow(\mib{r},E)$ and $N_\downarrow(\mib{r},E)$ is enhanced. 
This suppression of 
$\sqrt{N_\uparrow(\mib{r},E) N_\downarrow(\mib{r},E)}$ 
at $E \sim \Delta_0$ is associated 
with the fact that the suppression of $\kappa(\mib{r})$ near the vortex 
core becomes eminent in the $U/t=3.0$ case [Fig. \ref{fig:K-r}(d)] 
compared with the $U=0$ case [Fig. \ref{fig:K-r}(c)].

%%%%%%%%%%%%%%%%%%%%%%%%%%%%%
\begin{figure}[tbh]
\includegraphics[width=7cm]{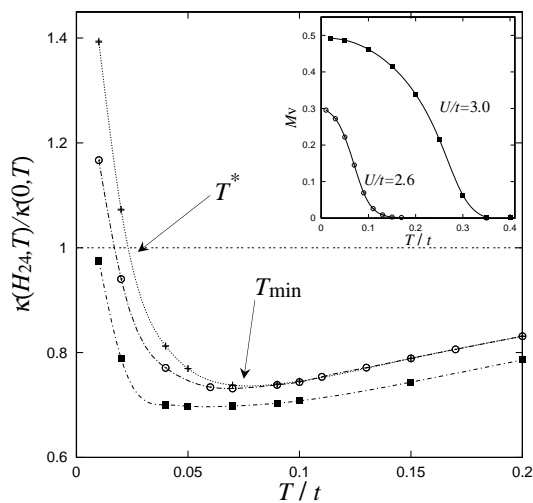}% Here is how to import EPS art
\caption{\label{fig:K-T}
$T$-dependence of normalized thermal conductivity 
$\kappa(H_{24},T)/\kappa(0,T)$.
$H_{N_r}=H_{24}(=0.001736)$, 
$U/t$=0.0($+$), 2.6($\bigcirc$) and 3.0($\blacksquare$).
The inset shows the $T$-dependence of the magnetic moment 
at the core site for $U/t=2.6$ and $U/t=3.0$.
}
\end{figure}
%%%%%%%%%%%%%%%%%%%%%%%%%%%%%

As mentioned before, there are two characteristic temperatures 
$T^{\ast}$ and $T_{\rm min}$ that govern the $\kappa(H)$ 
behavior in Fig. \ref{fig:K-H}.
$T^{\ast}$ is the temperature at which the $H$-dependence of 
$\kappa(H)$ is almost constant at low $H$ and  $T_{\rm min}$ is 
the ``reentrance'' temperature. 
In Fig. \ref{fig:K-T}, we show the $T$-dependence 
of $\kappa(H_{24},T)/\kappa(0,T)$ for the three $U$ values together with
the magnetic moment at the core. 
Since $T^{\ast}$ is the temperature corresponding to 
$\kappa(H)/\kappa(0)=1$, as $U$ increases, $T^{\ast}$ decreases 
and ultimately $T^{\ast}\rightarrow 0$ for $U/t=3.0$. 
This explains why there is no plateau curve observed 
until the very low $T$ in Fig. \ref{fig:K-H}(c).

As for the ``reentrant'' behavior in $\kappa(H)$, 
$T_{\rm min}$ occurs just above $T^{\ast}$. 
For $T^{\ast}<T<T_{\rm min}$, 
the slope of the decreasing $\kappa(H)$ at low $H$ 
becomes more steep as $T$ increases.
However for $T>T_{\rm min}$, the slope gradually becomes gentle.
This corresponds to the ``reentrant'' behavior 
in $\kappa(H)$  as shown in Fig. \ref{fig:K-H}(b). 
It is noted from Fig. \ref{fig:K-T} that as $T$ further increases, 
the two $\kappa(H_{24},T)/\kappa(0,T)$ curves for $U\neq 0$
merge with that for $U=0$ because the local moment 
vanishes gradually as shown in the inset of Fig. \ref{fig:K-T}.
It is also noted from Fig. \ref{fig:K-T} 
that $T_{\rm min}$ decreases as $U$ increases 
because the AFM pushes down 
the $\kappa(H_{24},T)/\kappa(0,T)$ curve as a whole.
As mentioned above, this precisely corresponds 
to the data obtained by Sun \textit{et al.}~\cite{sun}.

Let us examine the present results in a wide perspective. 
As mentioned above, the experimental evidence 
for the field-induced AFM local order is abundant.
Neutron experimental results 
on LSCO~\cite{yamada,lake1,lake2,khaykovich} 
are directly connected 
to the present thermal conductivity calculation. Namely, 
the observed enhancement of neutron signals under $H$ 
is due to the local AFM order which leads 
in turn to the suppression of the zero-energy density 
of states around a core. This ZEDOS otherwise piles up 
as $H$ increases. 
This suppression of ZEDOS itself has 
been known for some time in Bi2212~\cite{maggio,pan} 
and YBCO systems~\cite{renner}. 
Because of this suppression, $\kappa(H)$ 
decreases with $H$ for underdoped cases, while 
$\kappa(H)$ increases with $H$ for overdoped cases. 
According to Fig. 4 in ref. ~\citen{sun},
the reversion is noted to occur at the optimal doping $x=0.16$. 
This doping dependence coincides with 
the observed tendency in LSCO by neutron scattering 
experiments performed extensively for wider dopings 
under zero field~\cite{yamada2} 
where an elastic peak for static AFM vanishes at overdoping.
It is natural to expect that a magnetic 
field could strengthen this tendency~\cite{lake1,lake2}.

It should be noted that, 
in the mixed state, the $T$-dependence of $\kappa$ 
cannot be a simple power law $T^\alpha$ with an integer $\alpha$, 
that is, neither $\alpha=2$ for the $d$-wave zero-field case 
nor $\alpha=1$ for the high field limit because LDOS is spatially nonuniform 
(see ref. ~\citen{takigawa2} for details).

Thermal conductivity is one of the direct probes 
for observing the low-lying quasiparticles 
formed around a vortex core through its field and temperature dependences. 
The complex, yet rich $\kappa(H)$ behaviors observed 
by Sun \textit{et al.}~\cite{sun} 
and Hawthorn \textit{et al.} ~\cite{hawthorn} 
yield a vivid picture of the spatial spectral 
structure of low-energy quasiparticles 
through our analysis. 
Together with our previous analysis~\cite{takigawa2} 
of site-selective NMR experiments on YBCO~\cite{halperin} and Tl2201 
systems~\cite{kakuyanagi1,kakuyanagi2}, 
the present calculation which yields a qualitatively 
consistent explanation for $\kappa(H)/\kappa(0)$ 
unambiguously demonstrates that, 
(1) in underdoped LSCO, the local AFM order or stripe order, 
depending on the interaction strength $U$ or doping level, must be present. 
(2) In overdopings, the ordinary $d$-wave vortex core picture is applicable.

We thank Y. Ando, M. Tanatar, Y. Matsuda and T. Sakakibara for useful discussions, 
and Y. Ando for sending experimental data.

%%%%%%%%%%%%%  REFERENCES

\end{document}